\documentclass[showpacs,preprintnumbers,twocolumn]{revtex4}
\usepackage{amsmath,amssymb,pstricks,wasysym,stmaryrd,enumerate,graphicx}
\begin{document}

\title{Poincar\'e recurrence theorem and  the strong CP-problem}
\author{Alex C. Kalloniatis\footnote{alexander.kalloniatis@dsto.defence.gov.au}
\footnote{Present address: Defence Science and Technology Organisation,
1 Thynne St., Fern Hill Park, Bruce, ACT 2617, Australia.} }

\address{
Special Research Centre for the Subatomic Structure of Matter,
University of Adelaide,
South Australia 5005, Australia}

\author{Sergei N. Nedelko \footnote{nedelko@thsun1.jinr.ru}}
\address{ Bogoliubov Laboratory of Theoretical Physics, JINR,
141980 Dubna, Russia}

\date{\today}
\preprint{ADP-04-28/T610}
\begin{abstract}
The existence in the physical QCD vacuum of nonzero
gluon condensates, such as $\langle g^2F^2 \rangle$, 
requires dominance of gluon fields with finite mean action density. 
This naturally allows  any real number value for the unit 
``topological charge'' $q$ characterising
the fields approximating the gluon configurations 
which should dominate the QCD partition function. 
If $q$ is an irrational number then the critical values of the 
$\theta$ parameter for which CP is spontaneously broken are dense in 
$\mathbb{R}$, which provides for a mechanism of
resolving the strong CP problem  simultaneously
with a correct implementation of $U_{\rm A}(1)$ symmetry. 
We present an explicit realisation of this mechanism within a QCD motivated
domain model. Some model independent arguments are given that suggest 
the relevance of this mechanism also to genuine QCD.
\end{abstract}
\pacs{12.38.Aw 12.38.Lg 14.70.Dj 14.65.Bt 11.15.Tk}
\maketitle

\section{Introduction}

The functional space ${\cal F}_A$
of gluon fields $A$ with finite classical action 
(in the infinite volume limit),
\begin{eqnarray}
\label{action1}
\lim_{V\to\infty}S_V[A]<\infty,
\end{eqnarray}
can be divided into equivalence classes according to 
integer values of topological charge
\begin{eqnarray}
\label{inttopcharge}
\nu=Q[ A]=\frac{g^2}{32 \pi^2} \int d^4x F^a_{\mu \nu} {\tilde F}^a_{\mu \nu} .
\end{eqnarray}
Integer values originate from the purely topological 
properties of the gauge group: 
the above integral can be rewritten as a surface integral and, 
since at infinity all fields can be at most pure gauge configurations, 
the value of that integral is determined by the gauge group alone. 
According to these equivalence classes, the QCD partition function 
can be written in the form of a Fourier series
\begin{eqnarray}
\label{fourier}
Z(\theta)=\sum_{\nu=-\infty}^{\infty}e^{i \nu \theta}Z_\nu, 
\end{eqnarray}
where $Z_{\nu}$ is given by a functional integral over the fields belonging to 
the $\nu$-th sector of  ${\cal F}_A$.  The representation 
Eq.(\ref{fourier}) is characteristic of non-abelian gauge groups and 
appeals to  the existence of non-perturbative gluon field configurations, 
namely multi-instantons.  
As a matter of fact, from the very beginning
the above construction is based on a quasi-classical principle: 
the functional integral is assumed to be dominated by fields with 
minimal classical action, while fields with infinite action are 
excluded from consideration {\it ad hoc}.
However, as is ultimately required by hadron phenomenology, 
the physical QCD vacuum is characterised
by finite nonzero gluon condensates \cite{SVZ79}, 
among which the lowest order condensate can be related to 
a mean action density times the coupling constant squared
\cite{clarification},
\begin{eqnarray}
\label{meanaction}
\langle g^2F^a_{\mu \nu}(x) F^a_{\mu \nu}(x) \rangle 
= 4\lim_{V\to\infty}\langle g^2S_V\rangle/V\not=0,
\end{eqnarray}
which can be provided only by the  class of
configurations already excluded from Eq.(\ref{fourier}),
the field configurations with extensive (scaling with the volume $V$)
classical action.  
One faces the necessity of defining the QCD partition function as an integral 
over the space ${\cal F}_{\cal A}$ of gluon configurations $\cal A$  
which are allowed to have nonzero classical action density at space-time
infinity. For instance a suitable requirement for ${\cal F}_{\cal A}$ would be
\begin{eqnarray}
\label{newspace}
\lim_{x^2\to\infty}F^a_{\mu \nu}(x) F^a_{\mu \nu}(x)={\rm const},
\end{eqnarray}
with a nonzero constant on the RHS, which can be related to the 
trace anomaly of the QCD energy-momentum tensor \cite{Nie77,Mink81}.
Condition Eq.~(\ref{newspace}) produces no essential difficulties since the 
infinity arising from the volume of the system is simply a matter of 
normalisation of the functional integral, or put differently, of 
normalisation of the vacuum energy.  
The actual values of the gluon condensates and the constant on the RHS of
Eq.~(\ref{newspace}) are defined by the minima of the 
quantum effective action. 
This indicates that the properties of the QCD physical vacuum, 
which are encoded in the various condensates and 
are responsible for confinement and the mode of realisation
of chiral symmetry, are due to purely quantum effects.   
Consequently, the fields to be identified as dominating the QCD functional 
integral must not have vanishing 
action density at infinity, otherwise the main effect, the
existence of condensates, would be omitted.
In order to define the functional integral over fields subject to 
Eq.~(\ref{meanaction}) in an analytical approach
one can represent ${\cal A}=B+A$ with the background $B$  
from the class of fields dominating the integral 
and $A$ being small localised fluctuations in this background.  
The integral over fields $B$ has to be performed exactly, 
while fluctuations $A$ can be treated perturbatively. 
The candidate dominating fields $B^a_\mu(x)$ could be required to 
satisfy another condition for almost all $x\in {\mathbb{R}}^4$
\begin{eqnarray}
\label{background}
F^a_{\mu \nu}(x) F^a_{\mu \nu}(x)={\rm const},
\end{eqnarray}
which, and this is important, neither forbids space-time variation of the 
strength tensor and the fields $B^a_\mu(x)$
nor devalues the importance of topological (singular) pure gauge field 
defects of various dimensions
distributed throughout the entire Euclidean space-time \cite{Mink81,NK2001}.  

One can of course look for conditions other than 
Eqs.~(\ref{newspace})  and  (\ref{background}) to  
attempt to specify background fields $B$ such that they would not be 
assumed to fill space-time almost everywhere and would be described in the 
spirit of dilute instanton gas or liquid models \cite{Shur96}.  
However, despite diluteness, any 
superposition of an infinite number (in the infinite volume) of instantons and 
anti-instantons represents a field with infinite 
classical action, which can be different from the class fixed by 
Eq.~(\ref{background}), but for sure is absent in the space of 
integration $\cal F_A$ subject to Eq.~(\ref{action1}).

Eq.~(\ref{meanaction}) has an important consequence:
in the space of fields $\cal F_A$ with 
extensive classical action the functional $Q[{\cal  A}]$
can take any real value, rational or irrational, finite or infinite in the 
limit $V\to\infty$. This signals that
the equivalence classes fed into the representation Eq.(\ref{fourier}) 
do not exist in the space $\cal F_A$.
Nevertheless, the notion of the mean value of the modulus $|Q[{\cal  A}]|$
\begin{eqnarray}
\label{meatopcharge}
 Q =\lim_{V\to\infty}\langle  | Q_V[{\cal  A}]|\rangle,
\end{eqnarray}
is well-defined on $\cal F_A$ and, as is known from phenomenology and 
lattice QCD, is a nonzero constant, related to the topological 
susceptibility $\chi$ of the QCD vacuum \cite{largeNc}.  
In both cases of fields satisfying  Eq.~(\ref{background})
and dilute superpositions of instantons, $Q$ can be expressed in terms of 
a mean action density.
There is a crucial difference however between instanton based approximations 
of the fields with infinite action and backgrounds
(as in the domain model for instance~\cite{NK2001,NK2002,NK2004,NK20041}) 
based on Eq.~(\ref{background}). 
The literal use of instantons to approximate fields with infinite action 
assigns integer values of topological charge to the fields
from $\cal F_A$ and thus transports integer units $\nu$ of topological
charge from Eq.~(\ref{fourier}) into QCD with condensates  
where such integers are irrelevant. 
Instead, approximations based on Eq.~(\ref{background}) leave the 
freedom to have any real values for $Q$  
for genuine gluon configurations and unit "topological charge"
$q$ for the fields approximating them in a particular model of QCD vacuum.

It was realised a long time ago   
(see \cite{Polyakov}, and, for a recent review ~\cite{ShG}) 
that the strong CP-problem is a problem of the quasi-classical 
treatment of the QCD functional integral built into the representation 
Eq.~(\ref{fourier}), 
and that incorporation of the long-range gluon configurations 
responsible for confinement could potentially remove this problem. 
However to date no way has been found to achieve this
and maintain simultaneously the resolution of the $U_{\rm A }(1)$ problem
traditionally attributed to instantons.
Scenarios based on the existence of an additional boson,
the axion, have been suggested and have become canonical~(\cite{Axion,ShG}).

Nevertheless, as will be discussed below in detail on the basis of a 
QCD motivated  model,
and as has been proposed in \cite{Mink81} within the general 
model independent consideration,  
the strong CP and $U_{\rm A}(1)$ problems have 
a mechanism for simultaneous resolution within QCD. 
As is stressed in  \cite{Mink81},
the order of thermodynamic limit and the limit $\theta\to0$ are in general 
not interchangeable 
and thus independence with respect to $\theta$ of the infinite volume QCD
partition function of QCD 
does not automatically lead to vanishing  topological succeptibility  
$\chi=\lim_{\theta\to0}\lim_{V\to\infty}V^{-1}d^2Z_V(\theta)/d\theta^2$.

 The QCD motivated model we shall consider in this article is
the domain model \cite{NK2001,NK2002,NK2004,NK20041}.
For completeness and clear identification of the origin of the effect 
discussed in this article a comment about the choice of boundary conditions 
in the domain model is in order. The field $B$ is considered as a 
background field and is subject to the condition Eq.(\ref{background}).  
The system is considered in a large but finite volume $V$. 
The total volume is split into a large number of subvolumes -- domains.
The fluctuation quark fields
are defined by the bag-like conditions on the domain boundaries.
The characteristic size of a domain is much smaller than the characteristic 
size of the total volume. 
The thermodynamic limit is understood as the limit when the size of the 
total volume
goes to infinity together with the number of domains while the size of 
domains stays finite.
This construction is sufficient for 
defining and for practical calculation of the 
Euclidean functional integral for the partition function.  
It should be stressed that bag-like conditions are very different from the 
(quasi-)periodic conditions typical for formulations based on a 
compactification of ${\mathbb{R}}^4$
to the torus. The most important feature for our considerations
is that the spectrum $\lambda$ of the Dirac operator under bag-like 
boundary conditions is asymmetric with respect to $\lambda\to -\lambda$, while 
(quasi-)periodic boundary conditions lead to a symmetric spectrum. 
In particular, this asymmetry is responsible for the formation of the 
quark condensate and resolution of the
$U_A(1)$ problem in the model as is discussed in detail in \cite{NK2004}. 
Nevertheless, there is no contradiction between our results and 
canonical considerations based on torus-like compactifications of  
${\mathbb{R}}^4$~\cite{LeuSmil92}. 
These are just two different complementary statements of the problem.

Within the domain model  the  mechanism for resolving the strong
CP-problem is realised in the following way. 
In the presence of nonzero gluon condensates 
irrational values of $Q$ are permitted and
lead to a realisation of CP in which the set 
of critical values of the $\theta$ parameter 
for which the CP-breaking is spontaneous \cite{Das71} 
is {\it dense} in the interval $[-\pi,\pi]$. 
As a consequence, the infinite volume partition function with 
infinitesimally small quark masses
\begin{eqnarray*}
Z=\lim_{V\to \infty} Z_{V}(\theta)=\lim_{V\to \infty} Z_{V}(0)
\end{eqnarray*}
is independent of $\theta$, and
\begin{eqnarray*}
&&\lim_{V\to \infty} \langle {\mathbb E} \rangle_{V}^{\theta}\equiv\lim_{V\to \infty} \langle {\mathbb  E} \rangle_{V}^{\theta=0}, 
\\ 
 &&\lim_{V\to \infty} \langle {\mathbb O} \rangle_{V,\theta}\equiv 0,
\end{eqnarray*}
for any  CP-even  and  CP-odd operators   ${\mathbb E}$ and ${\mathbb O}$
respectively,
which resolves the problem of CP-violation. Simultaneously
one finds that topological succeptibility in QCD with massive quarks
\begin{eqnarray*}
\chi^{\rm QCD}= - \lim_{V\to \infty}\frac{1}{V}\frac{\partial^2}{\partial\theta^2}Z_V(\theta)\not=0,
\end{eqnarray*}
is nonzero, independent of $\theta$, and satisfies the anomalous Ward identity, which 
indicates a correct implementation of the $U_A(1)$ symmetry. In particular,
in the chiral limit the mass of the $\eta'$ is  
expressed via the topological succeptibility $\chi^{\rm YM}$ of  pure 
gluodynamics in agreement with the Witten-Veneziano formula.

If $Q$ takes a rational value then only a finite number of 
such critical points exist in the interval $\theta\in[-\pi,\pi]$,
which results in the standard strong CP problem.

To apply a mechanical analogy, an irrational value of  $Q$
leads to a picture which is reminiscent of ergodic motion resulting in a 
dense winding of a torus by a trajectory \cite{Arnold}, 
while motion over closed trajectories can be associated with rational $Q$.

This  article is devoted to an investigation of this mechanism
on the basis of the domain model of QCD 
vacuum~\cite{NK2001,NK2002,NK2004,NK20041}. 
We omit here all details related to the motivation,  
formulation and numerous results of the model given in the above papers,
apart from those which are specifically needed here. 
In the next two sections the $\theta$ dependence in the model 
under consideration is elucidated.  
The Witten-Veneziano mass formula,  Gell-Mann--Oakes--Renner relation, 
the anomalous Ward identity and $\eta\pi\pi$ decay are specifically 
discussed in section~\ref{massrel}.
The final section is devoted to explanation of the relation between  
the mechanism of the strong CP resolution in the domain model and  
the Poincar\'e recurrence theorem of ergodic theory
which justifies the title of the present paper.

\section{Theta dependence in the domain model}

The model is given in terms of the following partition function for
$N\to\infty$  domains of radius $R$
\begin{eqnarray}
{\cal Z} & = & {\cal N}\lim_{V,N\to\infty}\int\limits_{\Omega_{\alpha,\vec\beta}}d\alpha d\vec\beta
\prod\limits_{i=1}^N
\int\limits_{\Sigma}d\sigma_i
\int\limits_{{\cal F}_\psi^i}{\cal D}\psi^{(i)} {\cal D}\bar \psi^{(i)}
\nonumber \\
&&\times \int\limits_{{\cal F}^i_A} {\cal D}A^i 
\delta[D(\breve{ B}^{(i)})A^{(i)}]
\Delta_{\rm FP}[\breve{ B}^{(i)},A^{(i)}]
\nonumber \\
&&\times e^{
- S_{V_i}^{\rm QCD}
\left[A^{(i)}+{ B}^{(i)}
,\psi^{(i)},\bar\psi^{(i)}]-i\theta Q_{V_i}[A^{(i)}+{ B}^{(i)}]
\right]},
\label{partf}
\end{eqnarray}
where the functional spaces of integration
${\cal F}^i_A$ 
and ${\cal F}^i_\psi$  are specified by the boundary conditions  
$(x-z_i)^2=R^2$
\begin{eqnarray}
\label{bcs}
&&\breve n_i A^{(i)}(x)=0, 
\nonumber\\
&&i\!\not\!\eta_i(x) e^{i(\alpha+\beta^a\lambda^a/2)\gamma_5}\psi^{(i)}(x)=\psi^{(i)}(x),
\label{quarkbc} \\
&&\bar \psi^{(i)} e^{i(\alpha+\beta^a\lambda^a/2)\gamma_5} i\!\not\!\eta_i(x)=-\bar\psi^{(i)}(x).
\nonumber
\end{eqnarray}
Here $\breve n_i= n_i^a t^a$ with the 
generators $t^a$  of $SU_{\bf c}(3)$ in the adjoint representation,
the $\alpha$ and $\beta^a$ are random chiral angles 
associated with the chiral symmetry violating boundary condition
Eq.(\ref{quarkbc})  in the presence of $N_f$  ($a=1,\dots,N_f^2-1$) 
quark flavours. The $\lambda^a$ are $SU(N_f)$ flavour generators normalised
such that ${\rm{Tr}}(\lambda^a \lambda^b)=2 \delta^{ab}$.  
Averaging over the angles $\alpha, \ \beta^a$
is included in the partition function.
The thermodynamic limit assumes $V,N\to\infty$ but 
with the density $v^{-1}=N/V$ taken fixed and finite. The
partition function is formulated in a background field gauge
with respect to the domain mean field, which is approximated 
inside and on the boundaries of the domains by
a covariantly constant (anti-)self-dual gluon field with the 
field-strength tensor of the form
\begin{eqnarray*}
F^{a}_{\mu\nu}(x)
=
\sum_{j=1}^N n^{(j)a}B^{(j)}_{\mu\nu}\vartheta(1-(x-z_j)^2/R^2), 
\label{domainfieldstrength}
\end{eqnarray*}
with $B^{(j)}_{\mu\nu}B^{(j)}_{\mu\rho}=B^2\delta_{\nu\rho}$.
Here  $z_j^{\mu}$ are the positions of the centres of domains in 
Euclidean space.
The measure of integration $d\sigma_i$ over parameters characterising 
domains can be found in any of the papers~\cite{NK2001,NK2002,NK2004,NK20041}. 
The background field in Eq.(\ref{domainfieldstrength})
is designed to satisfy Eq.~(\ref{background}).  

In a sense, this model represents a ``step-function'' approximation to
a class of  fields with infinite action which are assumed to dominate the 
QCD partition function. The building blocks of this approximation are domains
with the same mean size $R$ and mean action density $B^2$,
whose values are determined from the phenomenological string tension
in \cite{NK2001}. 
Moreover as the field in domains is taken to be 
(anti)-self-dual, a mean absolute value $q$ of the integral of the 
topological charge density over the domain volume $v=\pi^2 R^4/2$ 
is attributed to a domain, and is expressed through the domain radius and 
field strength as $q=B^2R^4/16$. This quantity can take any real values
but for $B,R$ fixed as in \cite{NK2001} $q\approx0.15\dots$. 
It should be noted also that domain boundaries are assumed to be populated by  
pure gauge singularities which look like topological defects
(instantons, monopoles, vortices and domain walls) but only locally; 
globally no topologically conserved number can be associated with them. 
The role of these defects in the model is to generate boundary conditions 
for the fluctuations of gluon and quark fields $A$ and $\psi$ inside 
domains~\cite{NK2001}. 
In particular, at the location of pure gauge singularities 
$\partial V_i$ colour quark currents have to satisfy the condition
\begin{eqnarray*}
\bar\psi(x)\!\not\!\eta_i(x)t^a\psi(x)=0, \  x\in\partial V_i.
\end{eqnarray*}
with $\eta^\mu_i$ being a vector normal to the surface $\partial V_i$ 
of the $i-$th domain where the singularities are located.
This leads to the condition Eq.~(\ref{quarkbc}) with arbitrary 
parameters $\alpha, \ \beta^a$. Simultaneously it is natural to require that  
colourless quark currents like
\begin{eqnarray*}
\bar\psi(x) \lambda^a\psi(x),
\bar\psi(x)\gamma_5 \lambda^a\psi(x),
\bar\psi(x)\!\not\!\eta_i(x)\gamma_5 \lambda^a\psi(x), \ {\rm etc},
\label{continuity}
\end{eqnarray*}
are continuous at $\partial V_i$. 
Such continuity conditions Eq.(\ref{continuity})
fix the parameters $\alpha, \ \beta^a$ to be the same at 
all singular surfaces.

Apart from chiral angles and neglecting gluon fluctuations
$A$ in lowest order of the perturbation expansion, 
the integrations in the partition function Eq.~(\ref{partf}) give
\cite{NK2004}  
\begin{eqnarray}
Z_V(\theta)={\cal N} \int\limits_{\Omega_{\alpha\beta}}
d\alpha d\vec\beta \exp\{-V{\cal F}(\alpha,\beta,\theta)\},
\nonumber
\end{eqnarray}
with the free energy density as a function of chiral angles
in the presence of infinitesmally small quark masses $m_i$ 
\begin{eqnarray}
{\cal F}=-\frac{1}{v}
\ln[\cos  q(W_{N_f} -\theta)]  -\aleph \sum_{i=1}^{N_f} m_i\cos{\Phi_i},
\label{freeen}
\end{eqnarray}
and where 
\begin{eqnarray}
W_{N_f} = \sum_{i=1}^{N_f}{\rm arctan }({\rm tan}\Phi_i), \ \Phi_i=\alpha+B_i .
\nonumber
\end{eqnarray}
The functions $B_i$ for various numbers of flavours are
\begin{eqnarray} 
B_1 &=& 0 \ \ {\rm for} \ \ N_f=1,
\label{PhiNf1} \\
B_1&=&\frac{|\vec\beta|}{2}, \ \ B_2=-\frac{|\vec\beta|}{2}, 
\ \ {\rm for} \ \ N_f=2,
\label{PhiNf2}
\end{eqnarray} 
and, rather than give explicit representations 
for $N_f=3$, we give several identities useful for later purposes:
\begin{eqnarray}
\sum_{i=1}^3 B_i   &=& 0 \label{tracelessness}  \\ 
\sum_{i=1}^3 B_i^2 &=& \frac{1}{2} \sum_{a=1}^8 (\beta^a)^2 \\
\sum_{i=1}^3 B_i^3 &=& \frac{3}{4\sqrt{3}} \sum_{a=1}^3 (\beta^a)^2 \beta^8 
+ \dots 
\label{Nf3identities}
\end{eqnarray}
where the dots denote other cubic terms but involving
$a=4, \dots, 8$. Eqs.(\ref{PhiNf2}-\ref{Nf3identities})
come from the diagonalisation of the fermionic boundary conditions
Eq.(\ref{quarkbc}) in the flavour space.
The quantity $\aleph$
has been explicitly computed in~\cite{NK2004} 
and appears in the domain model as a function of 
the field strength $B$ and domain radius $R$
\begin{eqnarray*}
\aleph
&=&\frac{1}{\pi^2 R^3}
{\rm Tr}\sum_{k=1}^\infty \frac{k}{k+1}
\left[
M(1,k+2,z)
\right. \nonumber \\
&&\left. -\frac{z}{k+2}M(1,k+3,-z)-1
\right],
\end{eqnarray*}
where $M$ is the confluent hypergeometric function, 
$z=\hat n BR^2/2$  and the colour Tr denotes summation over 
elements of diagonal color matrix $\hat n$. 
With $B,R$ as determined in \cite{NK2001} $\aleph=(237.8 {\rm{MeV}})^3$.
It determines the value of the quark condensate 
\begin{eqnarray*}
\langle \bar\psi \psi\rangle = -\aleph
\end{eqnarray*}
and  arises from
the asymmetry of the Dirac operator spectrum \cite{DGS98} 
with the boundary conditions Eq.(\ref{quarkbc}).   
Due to Eq.(\ref{tracelessness}) (which applies for any $N_f$) we have
\begin{eqnarray*}  
W_{N_f} = N_f\alpha \, ({\rm mod} \ \pi).
\end{eqnarray*}
Thus the anomalous part of free energy density,
that involving the topological charge $q$, is independent of the $\beta^a$ 
which manifests the Abelian property of the anomaly. 
The significance of Eq.(\ref{freeen}) is that it
is derived directly from the domain {\it ansatz} 
for the QCD vacuum but also
corresponds to the zero momentum limit of the effective
chiral Lagrangian of \cite{largeNc} up to the somewhat
different encoding of the anomaly. A task in
the following is then verifying that this encoding
of the symmetries of low energy phenomenology reproduces the expected phenomena
of spontaneous chiral symmetry breaking and resolution of the $U_A(1)$
problem but more importantly that it also solves
the strong CP problem within this QCD vacuum approach.
  
In the thermodynamic limit $V\to\infty$ the minima of  ${\cal F}$
define the vacua of the system under consideration. 
For massless quarks $m_i\equiv0$  the anomalous 
term gives a discrete set of degenerate minima at
\begin{eqnarray}
\label{alkl1}
\alpha_{kl}(\theta)=
\frac{\theta}{N_f}+\frac{2\pi l}{qN_f}+\frac{\pi k}{N_f} \ \ (k\in Z, l\in Z).
\end{eqnarray}
Thus due to the anomaly 
the continuous $U_A(1)$ is reduced to a discrete group of  
chiral transformations connecting the minima Eq.(\ref{alkl1}) with each other. 
In the absence of additional sources violating chiral symmetries 
like quark masses, this discrete symmetry is sufficient to provide for 
zero mean values of all chirally non-invariant operators.

\section{Irrational value of parameter $q$}

There is a crucial difference between rational and irrational values of $q$. 
For $q$ rational there is a {\it finite} number of  distinct minima 
and all others are $2\pi$-equivalent to one of these minima. 
In other words, there is a periodic structure
which allows one to split the set Eq.~(\ref{alkl1}) into finite number of 
equivalence classes. 

If $q$ is an irrational number then all minima in the interval 
$\alpha\in[-\infty,\infty]$ are distinct: there are no minima which 
are $2\pi$-equivalent to each other. The distance between a pair of minima is
$$
\Delta(k_1,l_1|k_2,l_2)=2\pi m(k_1,l_1|k_2,l_2) +\delta(k_1,l_1|k_2,l_2)
$$
with $m(k_1,l_1|k_2,l_2)$ some integer, and 
$\delta(k_1,l_1|k_2,l_2)$ a number in the interval $(0,2\pi)$ which is  
different for any pair $(k_1,l_1|k_2,l_2)$, $l_1\not=l_2$. 
Unlike the case of rational $q$ where only several
numbers $\delta$ arise,
the set of all $\delta(k_1,l_1|k_2,l_2)$ is 
dense in the interval  $(0,2\pi)$ for irrational $q$.
The periodicity, typical for a rational $q$, is lost for irrational values of 
$q$  and the set of all minima cannot be split into $2\pi$-equivalence 
classes.  
However, these  discrete  minima of the free energy are separated by 
infinite energy barriers in the infinite volume limit, and no flavour 
singlet Goldstone modes are expected.

\begin{figure}[htb]
\includegraphics{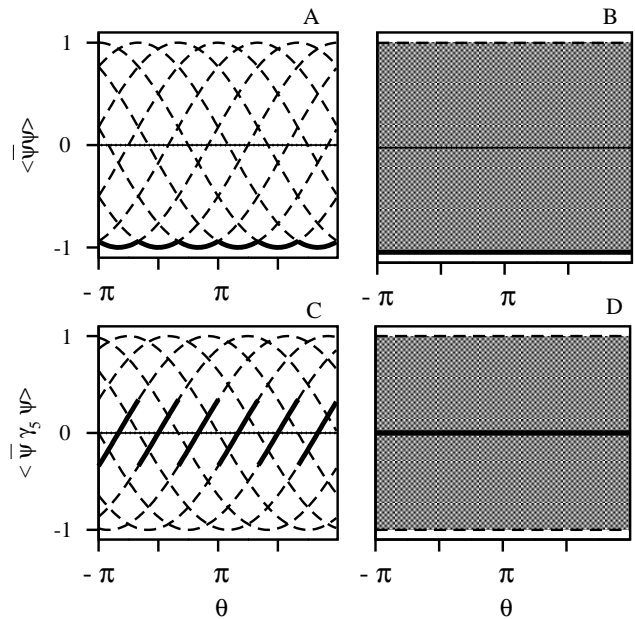}
\caption{The scalar (A and B) and pseudoscalar (C and D) quark condensates as  
functions of $\theta$ for $N_f=3$ in units of $\aleph$. 
The plots A and C are for rational  $q=0.15$,
while B and D correspond to any irrational $q$, for instance 
to $q=\frac{3}{2.02\pi^2}=0.15047\dots$
which is numerically only slightly different from $0.15$.
The dashed lines in A and C correspond to discrete minima of the free energy 
density which are degenerate for $m\equiv0$. The solid bold lines  
denote the minimum which is chosen by an infinitesimally small mass term for a 
given $\theta$. Points on the solid line in A, where two dashed lines cross 
each other, correspond to critical values of $\theta$, at which CP is broken 
spontaneously. This is signalled by the discontinuity in the pseudoscalar 
condensate in C.   For irrational $q$, as illustrated in B and D, 
the dashed lines densely cover the strip between $1$ and $-1$, and the set of 
critical values of $\theta$ turns out to be dense in ${\mathbb{R}}$. 
As will be discussed in the last section, 
an analogy can be seen between A (C) and motion on a torus over a closed 
trajectory, while a manifestation of the
Poincar\'e recurrence theorem resulting in dense winding of a torus 
(for example, see \cite{Arnold})  
can be recognised in B (D).} 
\label{fig:irrat}
\end{figure}

For any real $q$ this discrete symmetry and continuous 
non-singlet flavour chiral symmetry 
are spontaneously broken,
which becomes manifest if the infinitesimal masses are switched on.  
The term linear in masses  
in Eq.~(\ref{freeen}) selects a minimum 
for which $\alpha_{kl}(\theta)$ and the flavour non-singlet angles 
$\beta^a$ maximise $\sum_i^{N_f}m_i\cos(\Phi_i)$. There are specific values of 
the $\theta$ which are critical in the sense that  
two different minima $\alpha_{kl}(\theta_{\rm c})$ 
and $\alpha_{k'l'}(\theta_{\rm c})$ are degenerate in the presence of a 
mass term; these degenerate minima are CP-conjugates. 
The so-called Dashen phenomenon occurs \cite{Das71}: CP is broken 
spontaneously for  $\theta=\theta_{\rm c}$
while for non-critical values CP 
is explicitly broken.  
For rational $q$ there is only a finite number of $\theta_{\rm c}$ in the 
interval $(-\pi,\pi]$, 
which includes \cite{CrewNATO,largeNc} the value $\theta=\pi$. 
Irrational values of $q$ lead to a drastically different picture: 
the set of critical values $\theta_{\rm c}$ is dense in ${\mathbb{R}}$.    
In other words, any real value of $\theta$ is either critical or is a 
limit of a Cauchy sequence of critical values. 
CP is broken spontaneously 
in any arbitrarily small vicinity of any real value of $\theta$, 
in particular in the vicinity of $\theta=0$.  
Understood in the sense
of limiting values, the pseudoscalar condensate vanishes 
in both CP-conjugate vacua. 
The qualitative (and, in a sense, dramatic) difference 
in the consequences of $q$ being rational or irrational 
is illustrated in Fig.\ref{fig:irrat}.     

Note, that at $\theta=0$ CP is just an exact symmetry, it is not broken at all
(neither spontaneously, nor explicitly). 
The point $\theta=0$ is a limiting point 
of a set of critical points but it does not belong to this set:  
$\theta=0$ itself is not critical.  
Thus there is no contradiction with the Vafa-Witten theorem
\cite{Vafa1984}.

\section{Mass relations and anomalous Ward identity}
\label{massrel}

Flavour $SU_{\rm L}(N_f)\times SU_{\rm R}(N_f)$ is broken spontaneously 
and the appropriate number of corresponding Goldstone modes are expected, 
while $U_A(1)$ is realised in a more complex non-Goldstone form for any 
$\theta$, as shown in \cite{NK20041}.
To make this manifest at the level of the free energy Eq.~(\ref{freeen}) 
one can consider small variations of the variables $\alpha$ and $\beta^a$
in the vicinity of the minimising configurations  
$\{\alpha=\alpha_{\min}(\theta), B_1=B_2=B_3=0\}$. 
To this end it is sufficient to consider variations  
$\delta\alpha$, $\delta\beta^a$. 
We organise the expansion as follows:
\begin{equation}
{\cal F} = {\cal F}_0 + {\cal F}_{\beta} + {\cal F}_{\alpha}.  
\end{equation}
The leading term is 
${\cal F}_0= - N_f \aleph m  \cos[\alpha_{kl}(\theta)]$
which we have already identified
in \cite{NK20041} as generating the quark condensate
per flavour including its $\theta$ dependence,
$\langle {\bar q}_n q_n \rangle = - \aleph \cos[\alpha_{kl}(\theta)]$.
Fixing now $N_f=3$
we next consider the terms involving fluctuations in $\delta\beta^a$:
\begin{eqnarray}
{\cal F}_{\beta} &=& 
\frac{1}{4} \aleph m \cos[\alpha_{kl}(\theta)] \sum_{a=1}^{8}
(\delta\beta^a)^2 
\nonumber \\
&&- \frac{1}{8\sqrt{3}} \aleph m \sin[\alpha_{kl}(\theta)]
\sum_{a=1}^{3}
(\delta\beta^a)^2
\delta\beta^8 
\label{pionexpansion}
\end{eqnarray}
where we have showed all quadratic terms and one of the
cubic terms. Eqs.(\ref{tracelessness}-\ref{Nf3identities})
are used to derive this result. 
Let us now identify the flavour structure of the
fermionic boundary condition Eq.(\ref{quarkbc})
with that used in effective chiral theories \cite{largeNc} via
\begin{equation*}
i (\beta^a \lambda^a/2 + \alpha ) \equiv 
i \frac{\sqrt{2}}{F_{\pi}} (\pi^a \tau^a + \frac{1}{\sqrt{N_f}} \eta_0) 
\end{equation*}
where $\pi^a$ and $\eta_0$
are respectively identified as the non-singlet and singlet mesons
and the $\tau^a$ are normalised such that 
${\rm{Tr}}(\tau^a\tau^a)=\delta^{ab}$. 
The pion-decay constant is $F_{\pi}=93 \ {\rm MeV}$.
Thus we identify, for example for the neutral flavourless
mesons, the fluctuations as
\begin{eqnarray}
\delta \beta^3= \frac{2 \pi_0}{F_{\pi}},  \ 
\delta \beta^8= \frac{2 \pi_8}{F_{\pi}}, \  
\delta\alpha= \sqrt{\frac{2}{N_f}} \frac{\eta_0}{F_{\pi}}  
\label{mesonidentify}
\end{eqnarray}
Certainly what follows is 
not more than an illustration of what one can expect for
the meson spectrum and interactions:
a more solid investigation requires the construction of the effective meson
action and calculation of the relativistic bound state spectrum
in the spirit of \cite{NK2004}.
Nevertheless, if we identify the coefficients of the terms in
Eq.(\ref{pionexpansion}) which are quadratic in the
variations as ``meson'' mass terms then we obtain 
\begin{eqnarray*}
F_{\pi}^2m^2_{\pi/\eta}=2 m\aleph.
\end{eqnarray*}
This takes into account that for irrational $q$ one has
$\cos[\alpha_{kl}(\theta)]=1$ and  $\sin[\alpha_{kl}(\theta)]=0$
for any $\theta$
(see the bold solid lines in plots B and D in Fig.\ref{fig:irrat}).
Using our identification of $\aleph$ with the quark condensate, we 
thus recover the Gell-Mann--Oakes--Renner relation for masses of pions 
$(a=1,2,3)$ and the eta-meson $(a=8)$.

More significantly for the strong CP-problem
we observe that in higher order terms in the
decomposition those which are CP-even, namely involving
even powers of the fluctuations, come with a coefficient
$\cos[\alpha_{kl}(\theta)]=1$ for the allowed vacua.  
All CP-odd terms, those with odd powers of the fluctuations,   
are proportional to $\sin[\alpha_{kl}(\theta)]$ which exactly vanishes.
CP is thus unbroken. 
For example, the term which would correspond to an $\eta\pi\pi$ 
interaction is the cubic term in Eq.(\ref{pionexpansion}). 
Using the identifications Eq.(\ref{mesonidentify}), this is  
\begin{eqnarray}
-\sin[\alpha_{kl}(\theta)]\frac{m_{\pi}^2}{2 \sqrt{3}F_\pi}\pi_8(\pi_0)^2.
\label{etapipi}
\end{eqnarray}
For $q$ irrational this term exactly vanishes:
the decay $\eta\rightarrow\pi\pi$ is exactly suppressed.
For $q$ rational, we can expand Eq.(\ref{etapipi}) 
in the vicinity of $\theta=0$ using Eq.(\ref{alkl1}) 
and obtain $-\frac{\theta}{6\sqrt{3}}(m_{\pi}^2/F_{\pi}) \pi_8(\pi_0)^2$
which is  the analogue of Eq.~(8) in \cite{CdVVW79}
(and similar results elsewhere) for three degenerate quarks.  
In the context of approaches such as \cite{CdVVW79},
there is an {\it a priori} assumption that one can expand in small $\theta$: 
the $\theta$ dependence is first chirally rotated
from the topological charge density term into the quark
mass terms and then an expansion performed for small quark mass.
Using the small upper bound of the neutron dipole moment \cite{thetameasure}
such considerations lead to the expectation of an 
unusually small value for $\theta$, leading to the familiar 
fine-tuning problem. 
However, in the context of irrational topological charge
in the domain model, as a particular study of the
class of fields specified by the condition Eq.(\ref{background})
where the $\theta$ dependence can
be handled exactly, we see that such an expansion 
misses the possibility of the cancellation of CP-conjugate
vacua in CP-odd quantities. 
The absence of CP-violating hadronic processes 
such as $\eta\rightarrow \pi \pi$ (or, if baryons
could be included in these considerations, a neutron dipole moment) 
is not then indicative
of vanishing $\theta$; physical observables are independent
of $\theta$, as in Figs.\ref{fig:irrat}(B,D), and still CP symmetry
is exact in QCD with condensates, confinement
and the correct realisation of chiral symmetry, even in the presence of
the $\theta$ term.

Finally we turn to the fluctuations in the singlet directions $\alpha$.
Consider then the integral  
\begin{eqnarray*}
Z_V(\theta) = {\cal N} \int_{-\infty}^{+\infty}
 d\alpha \ e^{-V{\cal F}(\alpha,\vec\beta_{\rm min},\theta)}
\end{eqnarray*}
with $V=vN$ and finite $N\gg1$. 
The integral can be computed by means of the saddle-point approximation. 
The set of minima of the free energy density is obtained from the solution of equation
\begin{eqnarray*}
{\cal F}'_{\alpha}(\alpha,\vec\beta_{\rm min},\theta)&=& 
-v^{-1}\tan(q N_{f}\alpha-2qk\pi-q\theta-2l\pi )  \nonumber \\
&& -  N_{f} m \aleph \sin(\alpha)=0.
\end{eqnarray*}
The free energy density and its second derivative at the 
minima $\alpha_{kl}$  is  
\begin{eqnarray*}
{\cal F}&=&- \aleph m N_f \cos[\alpha_{kl}(\theta)]
\nonumber\\
{\cal F}''_{\alpha}&=& \chi^{YM} N_f^2 + 
\aleph m N_f \cos[\alpha_{kl}(\theta)]
\end{eqnarray*} 
where $\chi^{YM}=q^2/v=B^4R^4/128\pi^2$ is the topological susceptibility
in the absence of quarks which in the domain model was 
evaluated in \cite{NK2001} to be approximately $(197 \ {\rm{MeV}})^4$.
With the identification of $\delta\alpha$
of Eq.(\ref{mesonidentify}) we thus extract the Witten-Veneziano
mass formula for the eta-prime meson
\begin{equation}
m_{\eta'}^2 F_{\pi}^2 = 2N_f \chi^{YM} + m_{\pi}^2 F_{\pi}^2 
\end{equation}
in the physical vacuum.
Continuing with the evaluation of
the integral in the standard way we arrive at
\begin{eqnarray}
\label{zv}
Z_V(\theta) &=& \lim_{L\to\infty}{\cal N}_{L,N}
\nonumber\\
&\times&\sum_{l=-L}^L\sum_{k=1}^{N_{f}-1}
\frac{  e^{N N_{f}v m  \aleph \cos[\alpha_{kl}(\theta)]}}
{ [\chi^{YM} N_f^2 + \aleph m N_f \cos[\alpha_{kl}(\theta)]]^{1/2}}
\nonumber \\
&\times&[1+{\cal O}(m)] [1+{\cal O}(1/N)],
\label{partitionsum}
\end{eqnarray}
with $Z_V(0)=1.$ 
The sum above runs over all local minima of the free energy, 
enumerated by $k$, $l$.

Recall now that the topological susceptibility 
\begin{eqnarray}
\label{defchi}
\chi &\equiv& \left(\frac{g^2}{32\pi^2}\right)^2 \int d^4x 
\langle F\cdot\tilde{F}(x) F\cdot\tilde{F}(0)\rangle 
\nonumber\\ 
&=&-\lim_{V\to\infty}\lim_{\theta\to0} \frac{1}{Z_V(0)} \frac{1}{V}
{\frac{\partial^2}{\partial \theta^2}} Z_V(\theta).
\end{eqnarray}
It can thus be extracted from the partition function by 
the double variation with respect to $\theta$. 
It is important to note here that this relies on 
changing the order of taking the derivatives with respect to $\theta$ 
and functional integration, which is  well-defined
only for a certain regularized 
form of the functional integral.
In particular the total volume $V$ must be large but finite. 
In our context it means that  
the differentiation of Eq.(\ref{partitionsum}) 
with respect to $\theta$ must be performed  before taking the 
thermodynamic limit $N\to\infty$. When $V$ is large but finite 
$Z_V(\theta)$ depends on theta. After the second derivative 
over $\theta$ is taken the limit $N\to\infty$
chooses the global minimum, i.e. $l,k$ corresponding to the global minimum.
The key point here is that for irrational $q$,
$\forall \ \epsilon>0$ and $\forall \ \theta$ there exist 
integers $(l,k)$ (specifying the vacuum $\alpha_{kl}(\theta)$)  such that 
$1-\cos(\theta/N_f+ 2\pi l/q/N_{f})< \epsilon$. 
The existence of such a minimum eliminates 
dependence of observables on $\theta$. 
Moreover, taking the limit $\theta\to0$ is unnecessary in 
Eq.~(\ref{defchi}):  
as defined by Eq.~(\ref{defchi}) but with $\theta\to0$ omitted,  
 the CP-even quantity $\chi$ does not depend on $\theta$.
We repeat though that the most important point here is that  
differentiation with respect to $\theta$ and the thermodynamic limit
are not interchangable; 
 but deriving with respect to $\theta$ first is precisely how $\chi$ is 
obtained through Eq.~(\ref{defchi}).

Now let us use this to extract the topological susceptibility  in the presence
of light quarks. The aim, on the one hand, is to
check consistency with the anomalous Ward identity
of \cite{CrewRiv}, and on the other to confirm the
suppression of the susceptibility by quark loops,
as proposed by \cite{VenWit79} to fulfill the identities.
According to Eq.~(\ref{defchi}) the result is :
\begin{eqnarray}
\chi^{QCD}= \frac{m \aleph}{N_{f} }  + {\cal O}(m^2)      
\end{eqnarray}
and consistent (in Euclidean space) with the anomalous Ward identity of
\cite{CrewRiv}:
\begin{equation}
N_f^2 \chi^{QCD} = N_f m_{\pi}^2 F_{\pi}^2 + {\cal O}(m_{\pi}^4). 
\end{equation} 

\section{Discussion}
\subsection{The Poincar\'e recurrence theorem}
It is instructive to compare some aspects of the pictures 
arising for rational (in particular, integer)
{\it versus} irrational values of $q$. 
The fundamental difference between these two cases is clearly seen in 
Fig.~\ref{fig:irrat} but still requires precise  
formulation. As a matter of fact, plots (A,B) represent all inequivalent 
curves $-\cos[\alpha_{kl}(\theta)]$ while (C,D) show all inequivalent curves  
$\sin[\alpha_{kl}(\theta)]$. 

First of all, for rational $q$ there is a finite number 
$n_{\rm c}$ of critical points 
$(\{\theta^i_{\rm c}\}, \ i=1,\dots,n_{\rm c})$
in the interval $-\pi<\theta\le\pi$ and a countable set of critical points 
$(\{\theta^i_{\rm c}+2\pi j\}, \ i=1,\dots,n_{\rm c}, j\in \mathbb{Z})$  
in $\mathbb{R}$, distributed $(2\pi/n_c)$-periodically in $\mathbb{R}$.
In particular, $\theta=0$  is not critical  
while $\theta=\pi$ is critical for any rational $q$.  
The number of critical values $n_{\rm c}$ is determined by 
the integer numerator $q_1$
of a rational number $N_{f}q=q_1/q_2$.
For irrational $q$ the interval $-\pi<\theta\le\pi$ 
already contains a countable set of critical points  
$(\{\theta^i_{\rm c}\}, \ i=1,\dots,\infty)$ which is dense in this interval. 
Thus $\forall \theta\in \mathbb{R}$
 we have either a critical point or a limiting point of a 
sequence of critical points. In particular,
$\theta=0$ is such a limiting point for any irrational $q$. 

Furthermore, for  rational $q$ and $\forall\theta$,  
the set $\{{\cal C}_j(\theta)\}$ ($j=1,\dots, n_c$)
contains a finite number of points.
Here ${\cal C}_j(\theta)=\cos[\alpha_{j}(\theta)]$. 
The representation for $Z_V(\theta)$ is analogous to Eq.~(\ref{zv}) 
but contains a finite sum over $j=1,\dots, n_c$,
only one term of which dominates the sum $\forall \theta\not=\theta_{\rm c}$; 
the rest of the terms are exponentially suppressed at large volume.
Such a representation indicates that $Z_V(\theta)$
is differentiable to infinite order  $\forall \theta\not=\theta_{\rm c}$, 
which in particular means that CP-even observables can be computed for 
small $\theta$ as Taylor series in $\theta$. At critical $\theta$ 
the first derivative of $Z_V(\theta)$ is discontinuous.

Irrational values of $q$ change this behaviour  radically: 
${\cal C}=1$ is the limiting point of a sequence 
$\{{\cal C}_j(\theta)\}$ ($j=1,\dots, \infty$) $\forall\theta$.
This means by definition that $\forall\varepsilon>0$ and  
$\forall\theta$ $\exists J_\varepsilon(\theta)$ such that 
$1-{\cal C}_j(\theta)<\varepsilon$, $\forall j>J_\varepsilon(\theta)$. 
Simultaneously
$|{\cal S}_j(\theta)|<\varepsilon  $, $\forall j>J_\varepsilon(\theta)$. 
Here ${\cal S}_j(\theta)=\sin[\alpha_{j}(\theta)]$.  
The dependence of $J_\varepsilon$ on $\theta$ is discontinuous 
$\forall\theta$ and thus the sum in Eq.~(\ref{zv}) in the infinite volume 
limit represents the function $Z_V(\theta)$ which, strictly speaking, 
is not differentiable $\forall\theta$. 
However, for finite volume,
each term in the series is a differentiable function and the sum, 
regularised by $L<\infty$ and $N<\infty$,  
can be differentiated term by term, with a subsequent taking 
of the limits $V,L\to\infty$ which are well defined. 
In this sense Eq.~(\ref{zv}) gives a prescription for computing derivatives 
with respect to $\theta$. 

To  underscore that the character of the
$\theta$ dependence realised in the domain model 
does not represent anything unusual or ``exotic'' it is useful to map  
the above considerations to some textbook examples of application of  
the Poincar\'e recurrence theorem -- one of the 
corner stones of ergodic theory (for example, see~\cite{Arnold}).

\begin{figure}[htb]
\includegraphics[width=40mm,angle=0]{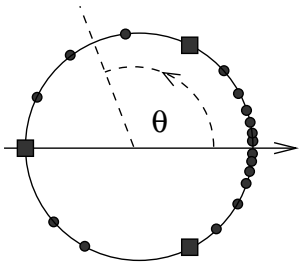}
\caption{Values of $\theta_l=g^l\theta$ as defined in Eq.~(\ref{gl}): 
boxes occur for rational $q=q_1/q_2$ with $q_1N_f=3$, while circles 
represent some of the points densely distributed, 
in particular, in the neighborhood of $\theta=0$ for irrational $q$. }
\label{circlefig}
\end{figure} 

The first straightforward example is just the present problem
slightly rephrased.
Consider the group of rotations   by angle $\beta$ acting on a point at  
polar angle $\theta$  on a circle.
The $l$-th element of this group corresponds to the angle 
\begin{eqnarray*}
\label{gl}
\theta_l=g^l\theta=\theta+\beta l \ ({\rm mod} \ 2\pi), \ \  \beta=2\pi b.
\end{eqnarray*}
According to the Poincar\'e recurrence theorem, 
for any rational $b$ there exists a number $l_b$ such that
$\theta_{l_b}=\theta$.  The
points represented in Fig.~\ref{circlefig} 
by the boxes correspond to the elements of the group. 
For any irrational $b$ the set $\{\theta_l, \ l=0\dots\infty\}$ is 
dense on the circle everywhere, particularly in the vicinity of $\theta=0$ 
as is illustrated by the filled circles in Fig.~\ref{circlefig}.
Identifying the $\theta$-parameter of QCD with the polar angle in  
Fig.~\ref{circlefig} and
the parameter $b$ with the ``topological charge'' $q$ by 
$b=1/N_{f}q$ establishes the desired map.

\begin{figure}[htb]
\includegraphics[width=80mm,angle=0]{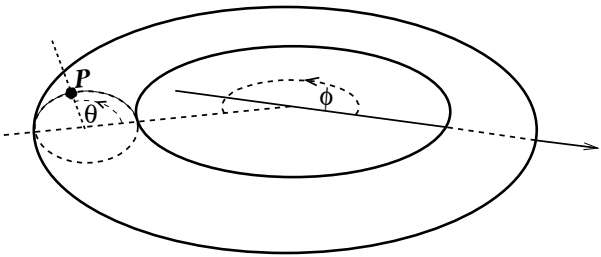}
\caption{Coordinates on a torus, as discussed in the text.}
\label{torusfig}
\end{figure}

Another simple example of the application of the Poincar\'e theorem  
is that of uniform motion on a torus,
\begin{eqnarray*}
\dot \theta=b_\theta,
\ \
\dot \phi=b_\phi,
\end{eqnarray*}
where $\theta$ and $\phi$ are the latitude and longitude of a point 
$P$ on the torus as is shown in Fig.~\ref{torusfig}.
If the ratio of velocities  $b=b_\theta/b_\phi$ is a
rational number then the point moves along a closed trajectory
which can be characterised by an integer winding number. 
In the case of irrational $b$ the trajectory
 is dense on the torus.  
If we take for simplicity $b_\phi=1$, identify $b_\theta=1/qN_{f}$
and replace the discrete parameter $2\pi l$ by a continuous variable $t$, 
then the plot $A$ in Fig.\ref{fig:irrat} would correspond to motion
on the torus over a closed trajectory with a winding number
equal to the number of critical points $n_{\rm c}$.  
An irrational $q$ leads to a correspondence between the curves densely 
covering the strip in plot B to the  
motion of a point on a torus over the trajectory 
densely covering the torus.

If we were to take the liberty of extrapolating 
the QCD motivated domain model considerations of this paper to genuine QCD, 
then we would conclude that for the class of background fields with 
$Q[{\cal A}]$ taking rational values the $U_A(1)$ problem can be solved but 
at the cost of simultaneously inviting the strong CP problem.
On the other hand,
irrational values of $Q[{\cal A}]$ are as natural for QCD with nonzero 
gluon condensates as rational (in particular, integer) values 
but providing clear signatures of resolving the $U_A(1)$ problem without 
creating the CP problem in strong interactions. 

\subsection{Model independent consideration} 
In this final part of the discussion we give arguments as to how the 
mechanism for resolving the strong CP-problem can be identified in  QCD 
in a model independent way. 
We start with the anomalous Ward identities of, for example, \cite{CrewNATO}.
These identities are model independent and arise from the symmetries of QCD 
and basic properties of the pseudoscalar meson spectrum.
A compact summary of this approach is given in Appendix B.1 of \cite{NK20041}.
Without repeating those derivations here, a key result obtained is
a differential equation
for the $\theta$ dependence of the matrix of condensates
\begin{equation}
V_{ij} \propto \langle 0| {\bar q}_i q_j |0\rangle
\end{equation}
whose phases are parametrised by angles $\phi_i$.
The true vacua are obtained by finding local minima  
with respect to perturbation of these angles by small $\omega_i$. 

From this one derives that a CP-even symmetry breaking term is
given by
\begin{equation}
\langle \epsilon H'_{even} \rangle = 2m \sum_i \cos \phi_i
\end{equation}
and a CP-odd symmetry breaking term is given by
\begin{equation}
\langle \epsilon H'_{odd} \rangle = 2m \sum_i \sin \phi_i.
\end{equation} 

One has  the following model independent requirements then:

{\it No $U(1)$ Goldstone boson,}  summarised in the equation
\begin{equation}
\frac{\partial}{\partial \theta} \sum_i \phi_i  = 1.
\label{nogoldstone}
\end{equation}

{\it Stationarity condition with respect to chiral perturbations,}
reflected by the equation
\begin{equation}
\sin \phi_i = {\rm independent \ of}\ i.
\label{statcond}
\end{equation}

{\it Minimum with respect to chiral perturbations},
\begin{equation}
\cos \phi_i > 0.
\label{minimum}
\end{equation}

{\it CP-conservation},
\begin{equation}
\sum_i \sin \phi_i = 0,
\label{CPinv}
\end{equation}
which might hold for only one value of $\theta$ (eg $\theta=0$)
or, ideally, may somehow hold for all $\theta$. 

$2\pi$-{\it periodicity}.
This is summarised in the equation
\begin{equation}
\sum_i \cos \phi_i|_{\theta} =  \sum_i \cos \phi_i|_{\theta+2n\pi} 
\label{2piperiod}
\end{equation} 
but cosine could just as well be replaced here by sine. This condition
emerges from previous conditions but only in the presence \cite{CrewNATO} of
the Dashen phenomenon at least at $\theta=\pi$. 

Each of these conditions is realised in the domain model \cite{NK20041}
and has appeared in previous sections of this paper. In the following
we shall only derive results from these equations.

From Eq.(\ref{nogoldstone}) we extract the result
$$
\sum_i \phi_i = \theta + c
$$
for some $\theta$-independent constant $ c$. 
We can give the individual angles as
$$
\phi_i = (\theta + c_i)/N, \  N^{-1} \sum_i c_i =  c.
$$
From Eq.(\ref{statcond}) we have in turn that
$$
\sin[(\theta+c_1)/N] = \sin[(\theta+c_2)/N]=\dots=\sin[(\theta+c_N)/N]
$$
thus  
$$
c_i = c\equiv \kappa\pi, 
$$
is independent of $i$ up to shifts of $2k\pi$.
All the allowed solutions to 
Eqs.(\ref{nogoldstone},\ref{statcond},\ref{minimum}) are classified
by
$$
\phi_1 =\dots=\phi_N= \varphi_{k} \equiv \varphi_0 +2 \pi k
$$
with the restriction $\varphi_0=(\kappa\pi+\theta)/N\in ]-\pi/2,\pi/2[$,
providing for positivity of the cosine.
At this point $\kappa$ is a real number.

Let us  first implement CP-invariance at $\theta=0$
just keeping $\kappa$ arbitrary. Since $\phi_i$ are
independent of $i$ it suffices that $\sin\phi_i=0$.  Thus
$$
\sin \kappa\pi/N = 0  
$$
which leads to  $\kappa=Nm.$
with $m$ integer.
Thus either $\kappa$ can be absorbed in $k$ or
should be set to zero. The latter choice leads to the statement 
in \cite{CrewNATO} that CP-invariance at
$\theta=0$ fixes unambiguously the integration constant 
from Eq.(\ref{nogoldstone}). 

Alternately  let us characterise $\kappa$ before seeking to 
impose CP-invariance at $\theta=0$. Note to this end that
any real number can be written as a product
of an integer and some number between zero and one,
\begin{equation}
\kappa=2l \xi, \ l\in \mathbb{Z}, \ \xi \in [0,1/2].
\end{equation}
This opens the possibility that different values of $l$ are appropriate 
for different ranges of $\theta$ subject to the above conditions. Thus far
for CP-invariance at $\theta=0$ we must have $l=0$. But for $\theta\neq 0$
nonzero values of $l$ 
satisfying condition
$$
2l\xi\pi+\theta=0 \ ({\rm mod} \ 2\pi).
$$
are now available due to the discrete nature of this
label.

We are now ready to map the model independent approach 
identically to the framework that 
emerged from the domain model via the 
the presence of the parameter $\xi$, which can have irrational values.
In view of Eq.(\ref{alkl1}) it is tempting to make
the identification
$$
\xi \equiv {1}/{q}.
$$
The conclusion we draw is that the spectrum of solutions to the theta
dependence of the QCD Ward identities is rich enough to exhibit the behaviour
seen in and originally derived in an explicit QCD motivated model, 
the domain model, which
solves the strong CP-problem. However the {\it interpretation} of the parameter
$\xi$ is possible at this stage only within the model, namely
as the mean value of a fraction of the topological charge  
related to the space-time regions where the dominating gluon configurations
in the QCD vacuum can be considered as homogeneous, or 
 succinctly -- the mean topological charge \textit{per} domain.

\section*{Acknowledgements}
ACK was supported by the Australian Research Council.
SNN was supported by the DFG, contract SM70/1-1
and partially by the grant RFBR~04-02-17370.
ACK thanks R.J. Crewther and B.-Y.~Park for helpful discussions.
SNN is grateful to G.~Efimov, A.~Isaev, V.~Papoyan, O.~Shevchenko and 
O.~Teryaev for clarifying discussions.


\begin{thebibliography}{9}
\bibitem{clarification}
It should be noted that the generally accepted notation
of Eq.(\ref{meanaction}) has to be understood in the sense of
$\langle: g^2 F^a_{\mu\nu}(x)F^a_{\mu\nu}(x): \rangle 
=4\lim_{V\to\infty}\langle: g^2 S_V:\rangle/V$, where the
normal ordering symbol $:\dots:$ means subtraction of
perturbative contributions, or more concisely normal
ordering with respect to the perturbative vacuum. Under this
prescription both sides of Eq.(\ref{meanaction}) are UV finite.
The UV finite and RG invariant quantity
$\langle:g^2 F^a_{\mu\nu}(x)F^a_{\mu\nu}(x): \rangle$
is the lowest gluon condensate of the OPE and QCD sum rules.
Usually however this ``normal ordering'' is implicit
and ommitted in notation for the gluon condensate.
\end{thebibliography}
\end{document}